\documentclass[a4,preprint,superscriptaddress,floatfix, showpacs]{revtex4}
\usepackage{amsmath}
\usepackage{graphicx}
\begin{document}
\author{A.A. Avetisyan}
\email{artak.avetisyan@ua.ac.be} %
\affiliation{Universiteit Antwerpen, Departement Fysica,
Groenenborgerlaan 171, B-2020 Antwerpen, Belgium}
\affiliation{Department of Physics, Yerevan State University, 1 A.
Manoogian, 0025 Yerevan, Armenia}
\author{B. Partoens}
\email{bart.partoens@ua.ac.be} %
\affiliation{Universiteit Antwerpen, Departement Fysica,
Groenenborgerlaan 171, B-2020 Antwerpen, Belgium}
\author{F. M. Peeters}
\email{francois.peeters@ua.ac.be} %
\affiliation{Universiteit Antwerpen, Departement Fysica,
Groenenborgerlaan 171, B-2020 Antwerpen, Belgium}
\affiliation{Departamento de F\'isica, Universidade Federal do
Cear\'a, Caixa Postal 6030, Campus do Pici, 60455-900 Fortaleza,
Cear\'a, Brazil}
\date{\today}
\title{Stacking Order dependent Electric Field tuning of the Band Gap in Graphene Multilayers}
\begin{abstract}
 The effect of different stacking order of graphene multilayers
on the electric field induced band gap is investigated. We
considered a positively charged top and a negatively charged back
gate in order to independently tune the band gap and the Fermi
energy of three and four layer graphene systems. A tight-binding
approach within a self-consistent Hartree approximation is used to
calculate the induced charges on the different graphene layers. We
found that the gap for trilayer graphene with the ABC stacking is
much larger than the corresponding gap for the ABA trilayer. Also
we predict that for four layers of graphene the energy gap
strongly depends on the choice of stacking, and we found that the
gap for the different types of stacking is much larger as compared
to the case of Bernal stacking. Trigonal warping changes the size
of the induced electronic gap by approximately 30\% for
intermediate and large values of the induced electron density.
\end{abstract}


\maketitle

\section{Introduction}

Graphene is a single layer of carbon atoms with hexagonal
symmetry~\cite{novoselov}. Multilayers of graphene can be stacked differently depending on the horizontal shift between consecutive graphene planes, leading to very different electronic properties~\cite{latil}, e.g. to various band structures.

A perpendicular electric field applied to bilayer graphene, with
the AB stacking, can open an electronic gap between the valence
and conduction bands~\cite{mccann}. This was shown indirectly by
transport measurements~\cite{castro, lieven}. Later on,
spectroscopic measurements confirmed the opening of a gap in the
energy spectrum ~\cite{ohta_Science, Li, zhang, kuzmenko,
zhang_Nature}. The extension of these bilayer results to three and
four layers of graphene was presented in Ref.~\onlinecite{Avetis}
in the case the perpendicular electric field was realized by a
single gate. It was found that such an electric field causes an
energy gap which was found to be a nonmonotonic function of the
gate voltage, and a reentrant opening and closing of the gap was
predicted as a function of the electric field strength. In
Ref.~\cite{koshino} the electronic band structure of the
ABA-stacked trilayer graphene in the presence of back and top
gates was invistigated.

 Recently, we generalized our previous results~\cite{Avetis} to the case when two,
i.e. top and back, gates were applied to three as well as to four
layers of graphene systems~\cite{Avetis2}. We found that due to
the trigonal warping the obtained results do not exhibit
electron-hole symmetry. A non-monotonic dependence of the true
energy gap in trilayer graphene on the charge density on the gates
was found. We also predicted an indirect gap with a non-monotonic
dependence on the gate voltage. Four layers of graphene exhibit a
larger energy gap as compared to the three layer system, which is
a consequence of the fact that Dirac fermions are present in the
AB stacked graphene multilayers in case of an odd number of
layers, while for an even number of stacked graphene layers only
charge carriers with a parabolic dispersion are present at low
energies~\cite{partoens7}.

 Using Raman spectroscopy measurements the graphitic flake thickness, i.e. the
number of graphene layers, can be obtained, as was demonstrated in
Refs.~\onlinecite{gutt} and~\onlinecite{graf}. In
Ref.~\onlinecite{gutt} a tunable three-layer graphene
single-electron transistor was experimentally realized showing a
transport gap near the charge neutrality point. To our knowledge,
up to now, no four layer system was studied experimentally.
Electrical tunable energy gap systems are of interest from a
fundamental point of view, but also for possible applications in
electronics (e.g. for transistors) and photonics (i.e. wavelength
tuning of a laser).

The electronic low-energy band structure of the ABC stacked
multilayer graphene was studied within an effective mass
approximation in Ref.~\onlinecite{koshino1}, with special
attention to the Lifshitz transition, in which the Fermi circle
breaks up into several pockets.

%Also, it is of great interest to study the behaviour of the
%induced gap in graphene multilayers for different types of
%stacking of the layers.
 In this paper
we study the effect of different ways of stacking of multilayers
of graphene on the electric field induced band gap by top and back
gates. We limit ourselves to those stackings that have been found
in graphite. The Bernal stacking (ABA), which has hexagonal
symmetry, is common and stable, but some parts of graphite can
also have rhombohedral one (the ABC stacking)~\cite{Lipson}. The
band structure of three and four layer graphene systems in the
presence of a perpendicular electric field is obtained using a
tight-binding approach, where we used a self-consistent Hartree
approximation to calculate the induced charges on the different
graphene layers. We found that the gap for trilayer graphene with
the ABC stacking is much larger than the one for the ABA stacking,
which was studied in Ref.~\onlinecite{Avetis2}.

 %and we found that it is easier to tune
%the Fermi energy in the forbidden gap in case of a $ABC$ stacked
%trilayer (for equal magnitude of charges on top and back gates).

Similarly for four layers of graphene the energy gap also strongly
depends on the choice of stacking, and is smallest in case of
Bernal stacking. When taking into account the circular asymmetry
of the spectrum, which is a consequence of the trigonal warping,
we found considerable changes in the size of the induced
electronic gap for the considered systems at intermediate and high
densities of total electrons induced on the layers.

This paper is organized as follows. A short overview of our
tight-binding approach with a description of the self-consistent
calculation are given in Sec.~\ref{sec2} for the ABC stacked three
layer graphene in the presence of top and back gates. The
corresponding numerical results are also discussed here. In
Sec.~\ref{sec3} we investigate four layer graphene with different
stacking order in the presence of top and bottom gates.
Sec.~\ref{sec4} summarizes our conclusions.

\section{Three layer graphene with the $ABC$ stacking in an external electric field} \label{sec2}

\begin{figure}
\centering
\includegraphics*[width=12cm]{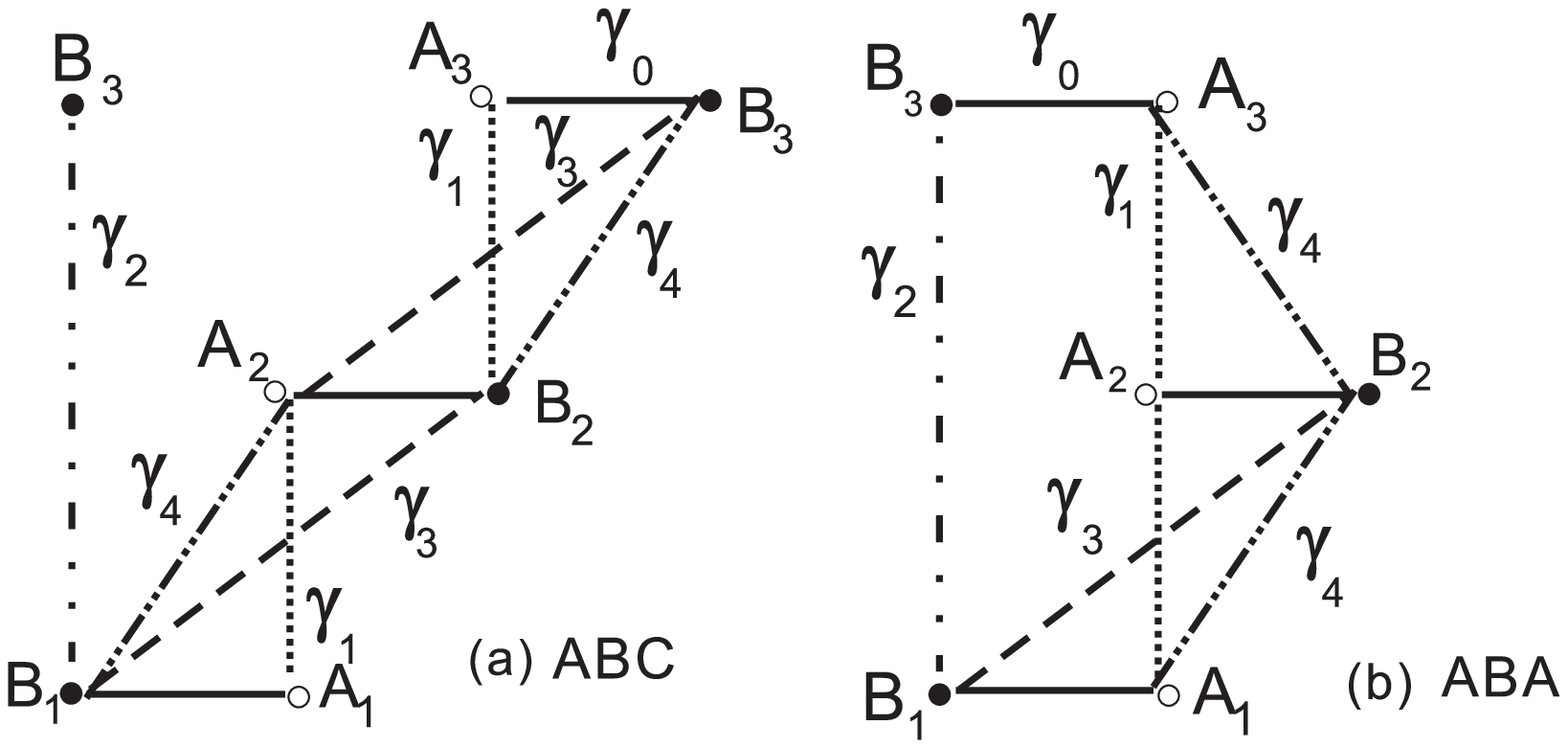}\caption{(Colour online)
Schematic of the different couplings between the sites for three
layers of graphene, where $A$-sites are indicated by white circles
and $B$-sites by black dots for: (a) the ABC stacking, and (b) the
ABA stacking.}\label{sch3layer}.
\end{figure}

We consider a system consisting of three layers of graphene with
the ABC stacking, which is modeled as three coupled hexagonal
lattices with inequivalent sites $A_{i}$ and $B_{i}$ ($i=1,2,3$ is
the layer number) with $A_{1}$ and $A_{2}$, as well as $A_{3}$ and
$B_{2}$ atoms on top of each other, as shown in
Fig.~\ref{sch3layer}. We use the Slonczewski-Weiss-McClure (SWMcC)
parameters, i.e. $\gamma_0, \gamma_1, \gamma_2, \gamma_3,
\gamma_4$ of tight-binding couplings for bulk graphite. Within
each layer the interaction between nearest neighbor $A_{i}$ and
$B_{i}$ atoms is described by the parameter $\gamma_0$. The strong
coupling between nearest layers, i.e. between $A_{1}-A_{2}$ and
$B_{2}-A_{3}$ atoms that lie directly above or below each other is
given by $\gamma_1$, and the weaker nearest layer coupling by
$\gamma_3$ ($\gamma_4$), i.e. between sites $B_{1}-B_{2}$ and
$A_{2}-B_{3}$ ($B_{1}-A_{2}$, $A_{1}-B_{2}$, $A_{2}-A_{3}$ and
$B_{2}-B_{3}$). The interaction between the next nearest layers
($B_{1}-B_{3}$) is determined by $\gamma_2$, as is shown in
Fig.~\ref{sch3layer}(a) and for comparison in
Fig.~\ref{sch3layer}(b) we show the unit cell for the ABA
trilayer. Using these parameters we compose the tight-binding
Hamiltonian for three layer graphene with the ABC stacking, which
has the form~\cite{guinea}

\begin{equation}
H=\left( \begin{array}{cc|cc|cc} D_{1} & H_{12} & H_{13} \\
H_{21} & D_{2} & H_{23}\\ \hline H_{31} & H_{32} & D_{3}\\
\end{array}\right),
\label{ham3lfree}
\end{equation}
where the rows and columns are ordered according to atom $A$ from
layer 1, atom $B$ from layer 1, atom $A$ from layer 2, atom $B$
from layer 2, etc, with the following two by two matrixes:
\stepcounter{equation}
\begin{equation}
D_{1}=\left( \begin{array}{cc} 0 & \gamma_0 f \\
 \gamma_0 f^* & 0
\end{array}\right),\quad
D_{2}=D^{\dag}_{1}, \tag{\theequation a} \label{ham3la}
\end{equation}

\begin{equation}
H_{12}=\left( \begin{array}{cc} \gamma_1  & -\gamma_4 f^* \\
 -\gamma_4f^* & \gamma_3 f
\end{array}\right),\quad
H_{21}=\left( \begin{array}{cc} \gamma_1 & -\gamma_4 f \\
 -\gamma_4 f & \gamma_3 f^*
\end{array}\right),
\tag{\theequation b}
 \label{ham3lb}
\end{equation}

\begin{equation}
H_{32}=H_{23}^{\dag}=\left( \begin{array}{cc} -\gamma_4 f   & \gamma_1\\
 \gamma_3 f^*& -\gamma_4 f
\end{array}\right), \quad
H_{31}=H_{13}=\left( \begin{array}{cc} 0 & 0\\
 0& \gamma_2/2
\end{array}\right),
\tag{\theequation c}
 D_{3}=D_{2}, \label{ham3lc}
\end{equation}
where
\begin{equation}
f(k_x,k_y) = e^{ik_xa_{0}/\sqrt{3}} + 2 e^{-ik_x
a_{0}/2\sqrt{3}}\cos{k_y a_{0}/2},
\end{equation}
with $a_{0}=2.46\AA$ the in-plane lattice vector length. The
Hamiltonian for the ABA stacking was discussed in
Ref.~\onlinecite{Avetis2}.

To control the density of electrons on the different graphene
layers and independently the Fermi energy of the system, a top
gate with a density of negative charges $n_{t}>0$ (the electron
excess density is positive) on it, and a back gate with a density
of positive charges $n_{b}<0$ are applied to the trilayer (a
schematic picture was presented in Fig. 1 of
Ref.~\onlinecite{Avetis2}). As a result a total excess density
$n=n_{1}+n_{2}+n_{3}$ is induced ($n=n_{t}+n_{b}$), with $n_{1}$
the excess density on the closest layer to the top gate, $n_{3}$
on the closest layer to the back gate, and $n_{2}$ is the excess
density on the middle layer. In our model the top or back gate
produces a uniform electric field $E_{t,b} =n_{t,b} e/ 2
\varepsilon_{0}\kappa$, and due to the induced charges on the
graphene layers, in its turn, create fields $E_{i}=n_{i} e/ 2
\varepsilon_{0}\kappa$  with $\varepsilon_{0}$ the permittivity of
vacuum and $\kappa$ the dielectric constant. There is a simple
relation between the charge density on the gates and the voltage
between the gate and the closest graphene layer: $V_{t,b}=
en_{t,b}d/2\varepsilon_0 \kappa$, where $d$ is the distance from
the gate to the closest graphene layer (usually $d$ is equal to
the oxide thickness, which is typically about 300nm). For our
numerical calculations we use the value $\kappa=2.3$, which
corresponds to graphene layers on $SiO_2$. The difference between
the charge densities induced on the individual layers of graphene
creates asymmetries between the first and the second layers, as
well as between the second and the third layers, which are
determined by the corresponding change in the potential energies
$\Delta_{1,2}$ and $\Delta_{2,3}$ \stepcounter{equation}
\begin{equation}
\Delta_{1,2}(n) = \alpha(n_{2}+n_{3}-|n_{b}|), \tag{\theequation
a} \label{del1a}
\end{equation}
\begin{equation}
\Delta_{2,3}(n) = \alpha (n_{3}-|n_{b}|), \tag{\theequation b}
\label{del1b}
\end{equation}
where $\alpha=e^{2}c_{0}/ \varepsilon_{0}\kappa$, with
$c_{0}=3.35\AA$ the inter-layer distance. The Hamiltonian
Eq.~(\ref{ham3lfree}) in the presence of the top and back gates is
modified and we have to add $\Delta_{1,2}(n)$, and
$-\Delta_{2,3}(n)$ to the first and third layer on-site elements
in Eq.~(\ref{ham3lfree}). The tight binding Hamiltonian operates
in the space of coefficients of the tight binding functions
$c(\overrightarrow{k})=(c_{A_{1}},c_{B_{1}},c_{A_{2}},c_{B_{2}},c_{A_{3}},c_{B_{3}})$,
where $c_{A_{i}}=c_{A_{i}}(\overrightarrow{k})$ and
$c_{B_{i}}=c_{B_{i}}(\overrightarrow{k})$ are the $i$-th layer
coefficients for $A$ and $B$ type of atoms, respectively. The
total eigenfunction of the system is then given by
\begin{equation}
\Psi_{\vec{k}}(\vec{r}) = \sum_{i=1}^{N_l} c_{A_i}
\psi_{\vec{k}}^{A_i} (\vec{r}) + \sum_{i=1}^{N_l} c_{B_i}
\psi_{\vec{k}}^{B_i} (\vec{r}), \label{coeff}
\end{equation}
with $N_l$ the number of layers. By diagonalizing the Hamiltonian
one can obtain the six coefficients (in Eq.~(\ref{coeff})) for
fixed values of the layer asymmetries, from which we obtain the
excess electronic densities on the individual layers:
\begin{equation}
n_{i} = \frac{2}{\pi}\int d k_{x} d k_{y} (|c_{A_i}|^{2}+|c_{B_i}|^{2}).
\label{exdens}
\end{equation}
The coefficients $c_{A_{i}}$ and $c_{B_{i}}$ depend on the
energetic band index. Here we are interested in the case when the
Fermi energy is located in the band gap, and in order to find the
redistribution of the electron density over the different layers
in the valence bands one should integrate Eq.~(\ref{exdens}) over
the Brillouin zone. The Fermi energy can be tuned into the opened
gap, when the magnitudes of the top and back gates are equal to
each other but with opposite charges on them. The other case when
the Fermi energy is located in the conduction or valence band was
discussed in Ref.~\onlinecite{Avetis2} for the ABA stacked
trilayer where we found that the obtained results do not exhibit
electron-hole symmetry in the presence of trigonal warping. Using
Eqs.~(\ref{ham3lfree}-\ref{del1b}) and~(\ref{exdens}) we evaluate
the energy gap $\Delta_{0}$ at the $K$-point and the true gap,
$\widetilde{\Delta}$, self-consistently for a fixed total density
$n_t+n_b=n_{1}+n_{2}+n_{3}$ (see Refs.~\onlinecite{mccann}
and~\onlinecite{Avetis2}).

In the following we will consider two cases. Firstly, we neglect
all interactions except between the nearest neighbour atoms in the
same layer and between the atoms of adjacent layers which are on
top of each other, i.e. we put
$\gamma_2=\gamma_3=\gamma_4=\gamma_5=0$. This leads to a circular
symmetric spectrum. In our calculations we used the parameter
$\gamma_0=3.12 eV$ which leads to an in-plane velocity
$\upsilon=\sqrt{3}\gamma_0 a/2\hbar\simeq10^6$ m/s, and for the
interlayer coupling strength, we take $\gamma_1=0.377$ eV (see
Ref.~\onlinecite{partoens}), and for the interlayer distance
$c_{0}=3.35\AA$. Secondly, the full interaction case is studied
where the interaction between the different atoms is expressed by
the SWMcC parameters ($\gamma_2=-0.0206,\gamma_3=0.29,
\gamma_4=0.12,\gamma_5=0.025$), i.e. the effect of warping is
included.

\begin{figure}
\centering
\includegraphics*[width=12cm]{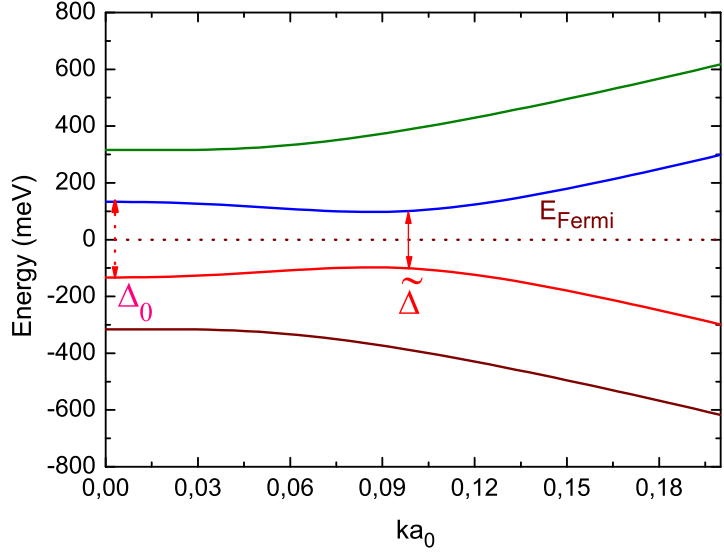}\caption{(Colour online)
The circular symmetric band structure of trilayer graphene with
the ABC stacking order around the $K$-point when charges on the
top and back gate are opposite but equal in magnitude, i.e.
$-n_b=n_t=10^{13}cm^{-2}$, for the case when only $\gamma_0,
\gamma_1 \neq0$. Horizontal dotted line is the Fermi level. The
true gap $\widetilde{\Delta}$ and the energy gap at the $K$-point
$\Delta_{0}$ are indicated.}\label{band2gtTril}
\end{figure}

 Fig.~\ref{band2gtTril} shows the band structure
for trilayer graphene with the ABC stacking when charges on the
top and back gates are opposite but equal in magnitude with
$-n_b=n_t=10^{13}cm^{-2}$ when only $\gamma_0, \gamma_1$ are taken
into account (with $\kappa=2.3$), and the Fermi energy is located
in the forbidden gap. Notice that there is conduction band -
valence band symmetry around the Fermi energy, and the true gap
$\widetilde{\Delta}$ occurs away from the $K$-point where the gap
is $\Delta_{0}=266meV>\widetilde{\Delta}=195meV$. For the ABA
stacking for the case when only $\gamma_0, \gamma_1\neq0$ the true
gap is zero for all densities.

\begin{figure}
\centering
\includegraphics*[width=12cm]{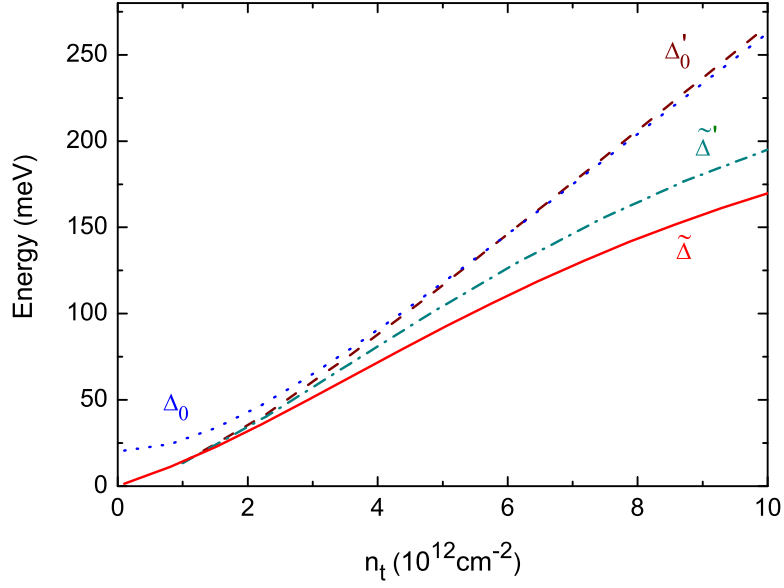}\caption{(Colour online)
The dependence of the gap $\Delta_{0}$ (dotted blue curve) at the
$K$-point, the true direct gap $\widetilde{\Delta}$ (solid red
curve) for the ABC trilayer graphene as a function of the top gate
density $n_{t}$ providing the back gate density is $-n_b=n_t$. For
comparison we show also the corresponding results, $\Delta'_{0}$
(dashed red curve) and $\widetilde{\Delta'}$ (dot-dashed blue
curve) when only $\gamma_0, \gamma_1\neq0$.}
 \label{delTril}
\end{figure}

When all the interactions between the different atoms are taken
into account the surface of constant energy is no longer circular.
In Fig.~\ref{delTril} we show the gap $\Delta_{0}$ at the
$K$-point (dotted blue curve), and the true direct gap
$\widetilde{\Delta}$ (solid red curve) for trilayer graphene with
the full interaction, as a function of the top gate density
$n_{t}$ providing the back gate density $-n_b=n_t$. For comparison
in the same figure we show also the corresponding results,
$\Delta'_{0}$ (dashed red curve) and $\widetilde{\Delta'}$
(dot-dashed blue curve) when only $\gamma_0, \gamma_1 \neq0$.
Notice, that for high densities ($-n_b=n_t \approx10^{13}cm^{-2}$)
the inclusion of the full interaction leads to a lowering of the
true gap by 30\%. It is interesting to note that similar values
for the energy gaps and the relative difference between them was
found for the case of bilayer AB graphene~\cite{Avetis2}: the true
gap for the AB bilayer at $-n_b=n_t \approx10^{13}cm^{-2}$ is
$142meV$ when $\kappa=2.3$ and $198meV$ for the case of
$\kappa=1$, when the full interaction is included. These results
compare with $169 meV$ ($\kappa=2.3$) and $207meV$ ($\kappa=1$)
for our ABC trilayer.

This similarity becomes more remarkable, if we compare the layer
densities induced by external gates for the ABA and ABC trilayers
with the AB bilayer. For the ABA trilayer, when only a back gate
was applied to the first layer~\cite{Avetis}, we found that
$n_1=6.1$, $n_2=3.2$ and $n_3=1.2$ at $n_b=10$ (in units
$10^{12}cm^{-2}$). The small amount of excess charges on the last
layers was explained by the fact that the graphene layers screen
the electric field and the layer asymmetries between the last
layers, counted from the gate, are very small. The true gap for
this system ($\widetilde{\Delta}=17 meV$) is smaller in comparison
with the bilayer case, where for the latter $n_1=8.3$ and
$n_2=2.8$ ($\widetilde{\Delta}=97.7meV$). Now, when only a back
gate is applied to the ABC trilayer we find that the densities on
the second and the third layers (counted from the back gate) are
very close to each other: $n_2=n_3\simeq2$ at $n_b=10$ and
$n_1=6.24$, which makes the ABC system distribution and the gap
(with $\widetilde{\Delta}=117meV$) similar to the AB bilayer ones.
In Fig.~\ref{sch3layer}(a) one can see that in the case of the ABC
stacking there are never 3 atoms stacked on top of each other, as
in the case for the ABA. As a result the electric field (of the
gate located near the first graphene layer for the ABC stacking)
penetrates easier to the last layers inducing excess charges,
while for the ABA stacking the electric field is much more
strongly screened.

\begin{figure}
\centering
\includegraphics*[width=12cm]{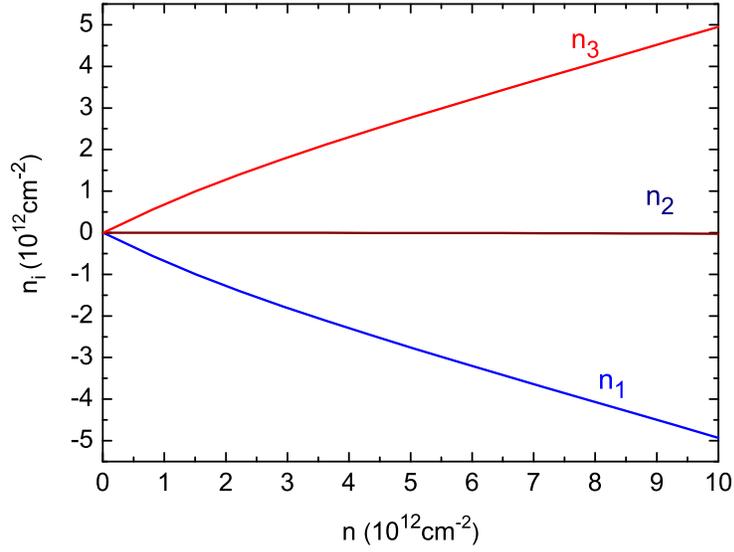}\caption{(Colour online)
The charge density $n_{i}$ on the different graphene layers for
the the ABC trilayer with $\kappa=2.3$ and with the full
interaction included, as a function of the charge density on the
top gate $n_{t}$ with the back gate density $n_b=-n_t$.}
\label{abc3lden}
\end{figure}

When both gates are applied to the ABC trilayer graphene (when the
full interaction is included) the excess charge densities at
$-n_b=n_t=10$, shown in Fig.~\ref{abc3lden}, on the outer layers
are $-n_1=n_3=4.9$ and in the middle layer is zero. Notice, that
the excess charge densities on the bottom and the top layers are
symmetric as in the case of the AB bilayer, as well as the gaps
have also similar values. While for the ABA trilayer it was
$n_1=-3.84$ and $n_3=3.67$ ~\cite{Avetis2}, and $n_2=0.17$ when
$-n_b=n_t=10$; the inclusion of the full interaction in the ABA
case makes the excess electron density in the middle layer
different from zero, and it opens a small gap about $5meV$. So, we
see that the ABC system has a large gap, comparable with the AB
bilayer one and behaves as a bilayer with shifted sheets, while
the ABA opens up much smaller gap and is similar to the case of an
AA stacked bilayer.
%Also, the charge
%densities on the outer layers, shown in Fig.~\ref{abc3lden} for the full
%interaction case, is very close to the densities induced in bilayer graphene with $AB$ stacking; for the $ABC$ trilayer at $n_b=-n_t=10$ in units %10^{12}cm^{-2}$ we found $n_2=-n_1=5$, while for bilayer graphene $n_2=-n_1=5.6$.
%gates it is no longer zero.

\begin{figure}
\centering
\includegraphics*[width=12cm]{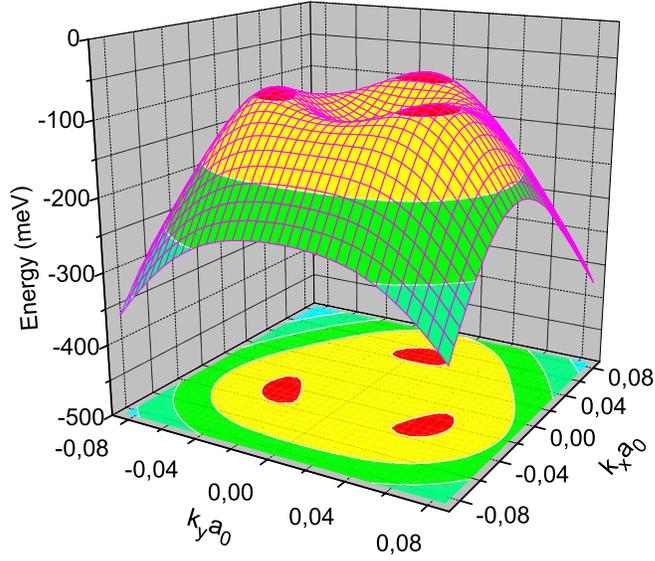}
\caption{(Colour online) The highest valence band, with the
corresponding contourplots for the ABC stacked trilayer graphene
near the $K$-point ($K$-point is chosen as the origin) with equal
but opposite charges on the top and back gate when
$n_{t}=-n_b=10^{13}cm^{-2}$. The Fermi energy is located in the
energy gap at $E=0$.}\label{3Dband3L}
\end{figure}

In the case of the previous studied ABA trilayer~\cite{Avetis} the
inclusion of trigonal warping leads to a non-monotonic behaviour
of the gaps as a function of gate voltage, as well as a much
stronger lowering of the true gap. Here, we found that the energy
gaps for the ABC stacked trilayer is much larger as compared to
the case of the ABA trilayer.
 Fig.~\ref{3Dband3L} shows a $3D$-plot and the
corresponding contourplot of the highest valence band for three
layer ABC stacked graphene near the $K$-point ($K$-point is chosen
as the origin, $\kappa=2.3$) for $n_{t}=-n_b=10^{13}cm^{-2}$. The
lowest conduction band is again symmetric with the highest valence
band just as in the case when only $\gamma_0, \gamma_1\neq 0$.
Here, for the ABC stacking we find three maxima, but did not find
additional maxima as in the case of the ABA
stacking~\cite{Avetis2} and as a result we do not observe an
indirect gap.

\section{Four layer graphene system in an external electric field} \label{sec3}

\begin{figure}
\centering
\includegraphics*[width=15cm]{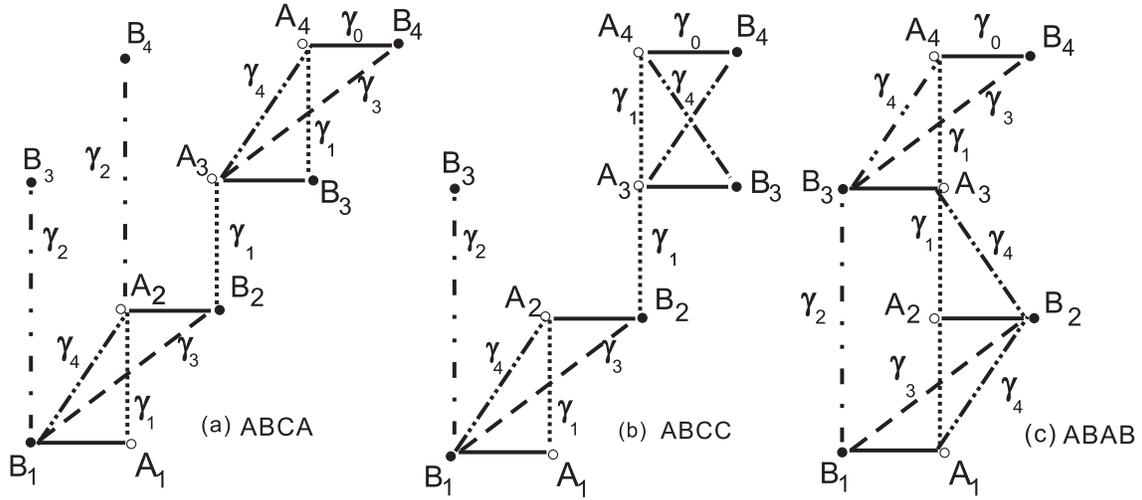}\caption{(Colour online)
Schematic of the couplings between the different ($A$-white and
$B$- black dots) sites  for four layers of graphene for: (a) the
ABCA, (b) the ABCC stacking, and (c) the ABAB Bernal
stacking.}\label{sch4layer}
\end{figure}

Now, we consider the four layer graphene system, which can be
arranged in many different ways as schematically shown in
Figs.~\ref{sch4layer}(a-c). The tight-binding parameters
$\gamma_i$ and the interaction between the individual carbon atoms
for all these cases are indicated in these figures. Four layer
graphene is described by the Hamiltonian
\begin{equation}
H=\left( \begin{array}{cc cc cc cc} D_{1} & H_{12} & H_{13} & H_{14}\\
H_{21} & D_{2} & H_{23} & H_{24}\\ H_{31} & H_{32} & D_{3} & H_{34}\\
H_{41} & H_{42} & H_{43} & D_{4}\\
\end{array}\right).
\label{ham4lfree}
\end{equation}
where $H_{ij}$ and $D_i$ with $i=1,2,3$ are the matrix elements of
the ABC trilayer given by Eqs.~(\ref{ham3la}-\ref{ham3lb}) and for
the ABCA stacking we have
 \stepcounter{equation}

\begin{equation}
H_{14}=H_{41}=\left( \begin{array}{cc} 0  &  0 \\
 0 &  0
\end{array}\right),\quad
H_{42}=H_{24}^{\dag}=\left( \begin{array}{cc} 0  &  0 \\
\gamma_2/2 &  0
\end{array}\right), \tag{\theequation a} \label{ham4lABCDa}
\end{equation}

\begin{equation}
H_{43}=H_{34}^{\dag}=H_{32}, \quad D_{4}=D_{3}, \tag{\theequation
b} \label{ham4lABCDb}
\end{equation}
while for the ABCC stacking these matrixes have the following
form:

\stepcounter{equation}

\begin{equation}
H_{14}=H_{41}=\left( \begin{array}{cc} 0  &  0 \\
 0 &  0
\end{array}\right), \quad
H_{42}=H_{24}^{\dag}=\left( \begin{array}{cc} 0  &  \gamma_5/2 \\
0 &  0
\end{array}\right),
\tag{\theequation a} \label{ham4layerABCCa}
\end{equation}

\begin{equation}
H_{43}=H_{34}^{\dag}=\left( \begin{array}{cc} \gamma_1  &  -\gamma_4 f \\
-\gamma_4 f  & \gamma_1
\end{array}\right), \quad
D_{4}=D_{3}. \tag{\theequation b} \label{ham4lABCCb}
\end{equation}
We consider a four layer graphene system with top and back gates,
which induce a total excess density $n=n_{1}+n_{2}+n_{3}+n_{4}$,
where $n_{i}$ is the excess density on the $i$th layer as counted
from the top gate. The corresponding change in the potential
energy between consecutive layers is

\stepcounter{equation}

\begin{equation}
\Delta_{1,2}(n) = \alpha(n_{2}+n_{3}+n_{4}-|n_{b}|),
\tag{\theequation a} \label{del4a}
\end{equation}

\begin{equation}
\Delta_{2,3}(n) = \alpha(n_{3}+n_{4}-|n_{b}|), \tag{\theequation
b} \label{del4b}
\end{equation}

\begin{equation}
 \Delta_{3,4}(n) =
\alpha( n_{4}-|n_{b}|). \tag{\theequation c} \label{del4c}
\end{equation}
By adding $\Delta^{II}=\Delta_{1,2}(n)$,
$\Delta^{III}=\Delta_{1,2}(n)+\Delta_{2,3}(n)$ and
$\Delta^{IV}=\Delta_{1,2}(n)+\Delta_{2,3}(n)+\Delta_{3,4}(n)$ to
the on-site  elements of the $II$, $III$ and $IV$ layer of the
ABCA or the ABCC four layer Hamiltonian, respectively, we obtain
the Hamiltonian in the presence of top and bottom gates. The eight
coefficients $c_{A_{i}}=c_{A_{i}}(\overrightarrow{k})$ and
$c_{B_{i}}=c_{B_{i}}(\overrightarrow{k})$, for fixed values of the
layer asymmetries defined by Eqs.~(\ref{del4a}-\ref{del4c}), can
be obtained by diagonalizing the corresponding Hamiltonian. The
electronic densities on the individual layers are given by
Eq.~(\ref{exdens}). The gaps $\Delta_{0}$, $\widetilde{\Delta}$
are evaluated self-consistently analogously as was done for the
three layer system.

\begin{figure}
\centering
\includegraphics*[width=12cm]{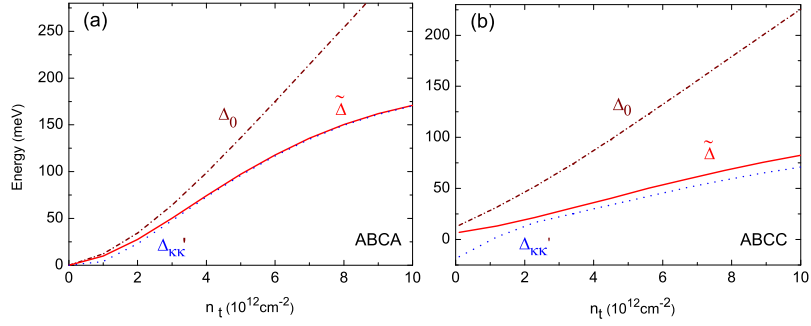}\caption{(Colour online)
The dependence of the gap $\Delta_{0}$ at the $K$-point
(dot-dashed red curve), the true direct gap $\widetilde{\Delta}$
(solid red curve) and the true indirect gap (dotted blue curve)
$\Delta_{kk'}$ as a function of the top gate density $n_{t}$ for
four layer graphene where we included the full interaction. The
back gate density $-n_b=n_t$ is the same (but opposite in sign) as
the top gate. Results are shown for: a) the ABCA stacking, and b)
the ABCC stacking.}
 \label{del4vt}
\end{figure}

\begin{figure}
\centering
\includegraphics*[width=12cm]{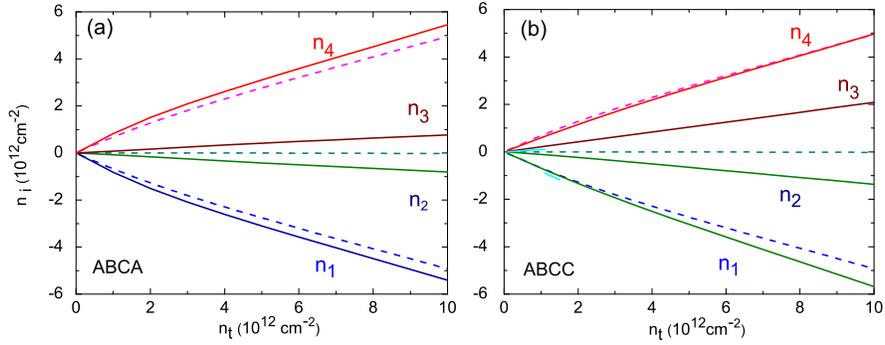}\caption{(Colour online)
The layer densities $n_{i}$ (solid curves) for the four layer
system  as a function of the charge density on the top gate
$n_{t}$ (providing $-n_b=n_t$) when the full interaction is
included: a) for the ABCA stacking and b) for the ABCC stacking.
In both cases we added the results for the layer densities
$n_{i}^{'}$ (dashed curves) for the ABC stacked trilayer when the
full interaction is included.}\label{4ldens}
\end{figure}

The variation of the gap $\Delta_{0}$ at the $K$-point (dot-dashed
red curve), the true direct gap $\widetilde{\Delta}$ (solid red
curve) and the true indirect gap (dotted blue curve)
$\Delta_{kk'}$ with the top gate density $n_{t}$ ($n_b=-n_t$) for
four layer graphene with the full interaction is shown in
Fig.~\ref{del4vt}(a) for the ABCA stacked four layer graphene  and
in Fig.~\ref{del4vt}(b) for the ABCC stacking. One can see that
for the ABCA stacking with full interaction the true direct gap is
very close to the corresponding gap in the case of a trilayer with
the ABC stacking, e.g., for $n_{t}=-n_b=10^{13}cm^{-2}$ the true
gap is about $171meV$ for four layer graphene with the ABCA
stacking and for the ABC trilayer it is $169meV$. In
Figs.~\ref{4ldens}(a) and~\ref{4ldens}(b) we present the layer
densities for the ABCA and ABCC four layer graphene systems,
respectively, and we include the curves for the densities in the
ABC stacked trilayer for comparison in both figures (dashed
curves). It is remarkable that the excess densities for the ABCA
system on the outer as well as on the inner layers are symmetric.
Notice, that the densities, shown in Fig.~\ref{4ldens}(a), on the
outer layers for the ABCA are very close to the ABC trilayer
graphene ones for all the values of $n_{t}$: at $-n_b=n_t=10$ for
the ABCA $n_4=-n_1=5.4$, while for the ABC trilayer graphene
$n_3=-n_1=4.9$ (in the units $10^{12}cm^{-2}$). We see that also
the ABCA four layer graphene behaves as the AB bilayer.
 The localization of the atoms (see Fig.~\ref{sch4layer})
 can explain why the excess electron densities on the outer sheets for the ABCA system are larger
 than the densities for the ABC trilayer, and even in comparison with the AB bilayer densities
 (which has $n_2=-n_1=3.7$ at
$-n_b=n_t=10$). For the ABCA system there are never 4 atoms on top
of each other, as is the case for the ABAB stacking. As a
consequence, the electric field of the top gate, e.g. at $A_{3}$
(see Fig.~\ref{sch4layer}(a)) is screened only by the $B_{2}$
atom. Similarly, the field of the back gate at $B_{2}$ is screened
only by $A_{3}$ atom. As a result, both these atoms feel the field
of the top as well as the back gate, which leads to a decrease of
excess charges on the inner layers (i.e. to a neutralization of
these charges by the opposite gates). However, an outer layer (of
the ABCA system) which is mainly charged by its closest gate does
not feel the further located gate, since the latter is screened by
the inner layers. In the AB bilayer the two sheets feel both
gates, and consequently the excess charges (by absolute value) are
less than in the outer layers of the ABCA. Due to this, also the
gap for the AB bilayer is less (see the gap value in previous
Section) than the ABCA one (for the same strength of the top and
back gates). We see also, that the gap is large when the amount of
excess charges in the inner layers is small (as it is for the ABCA
system).
 We found also that for
$\kappa=1$ the true gap is $183meV$ for the ABCA; the relative
difference with the case of $\kappa=2.3$ is only about 5\%.

In contrast, for the ABCC four layer graphene the excess density
on the third layer is larger, and on the fourth layer is smaller
than the corresponding densities found in the case of the ABCA
system. So, the increase in the excess densities as well as the
density asymmetry in the inner sheets leads to a decrease of the
gap. Also, the fact that the third and fourth sheets are not
shifted, i.e. they have the AA stacking order, explain that in the
ABCC four layer graphene the electric field opens up a smaller
gap. In both cases we found a much larger gap (about $170meV$ for
the ABCA stacking and $70meV$ for the ABCC stacking at
$n_{t}=-n_b=10^{13}cm^{-2}$) than in the case of the ABAB stacked
four layer graphene~\cite{Avetis2} (with $5meV$ for the same
density). So, we see that from all the systems, considered in this
paper and in Ref.~\onlinecite{Avetis2}, the Bernal stacking leads
to the smallest gap.

\begin{figure}
\centering
\includegraphics*[width=15cm]{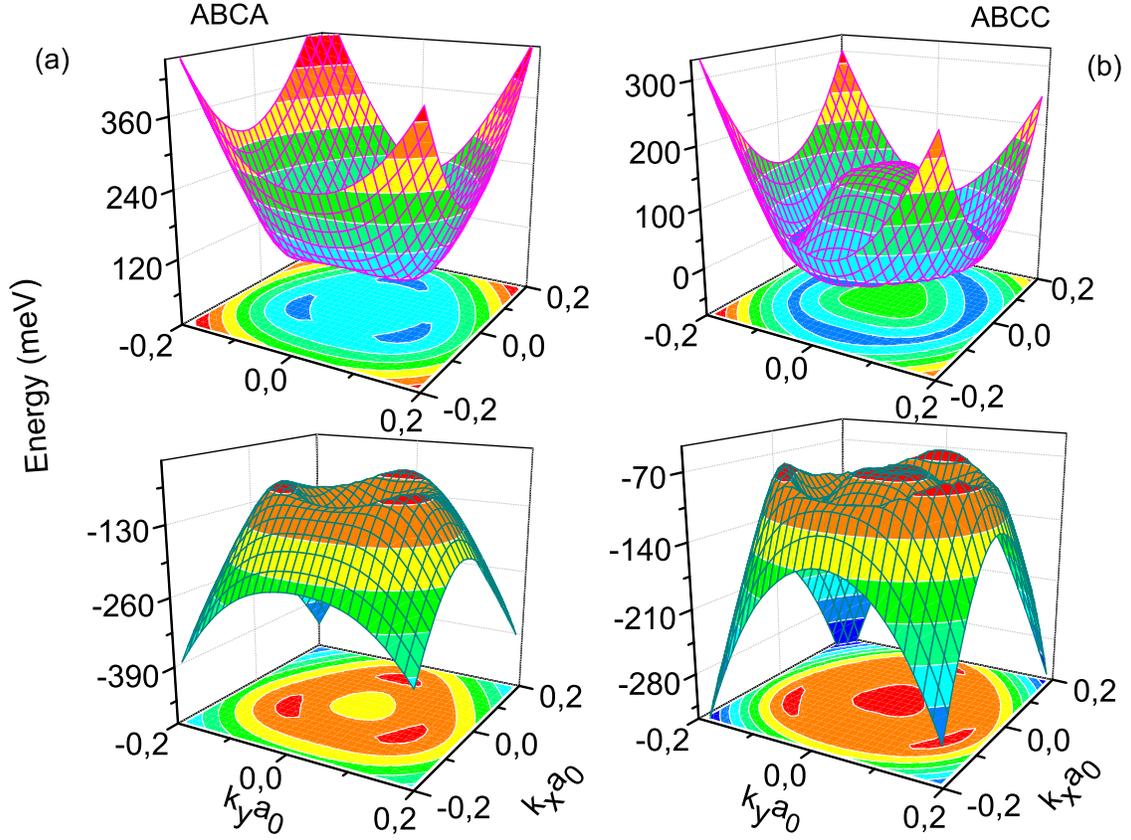}\caption{(Colour online)
$3D$-plots and corresponding contourplots of the highest valence
band (bottom figures) and the lowest conduction band (top figures)
around the $K$-point ($K$-point is chosen as the origin
$\kappa=2.3$) when $n_{t}=-n_b=10^{13}cm^{-2}$ for: (a) the ABCA
and (b) the ABCC stacking of four layers of
graphene.}\label{3Dabca}
\end{figure}

Figs.~\ref{3Dabca}(a) and (b) show $3D$-plots and corresponding
contourplots of the highest valence and the lowest conduction
bands near the $K$-point ($K$-point is chosen as the origin,
$\kappa=2.3$) in the case of $n_{t}=-n_b=10^{13}cm^{-2}$, for the
ABCA and the ABCC stacking, respectively. The conduction band for
the ABCC stacking has a "Mexican hat" shape maxima and minima on a
ring, as shown in the contourplot, e.g. there is a minimum at
$k_{x}a_{0}\simeq-0.17$ and $k_{y}a_{0}=0$. In its turn the
valence band has a local minimum between the two maxima at the
plane $k_{x}a_{0}\simeq-0.17$. The asymmetry between the
contourplots for the conduction and the valence bands for the ABCC
(see Fig.~\ref{3Dabca}(b)) leads to an indirect true gap. At low
densities there is a true direct gap for the ABCC, but due to the
overlap between the bands at different points in $k$-space the
indirect gap is negative as is shown in Fig.~\ref{del4vt}(b), i.e.
we have a semi-metal for low gate densities. For the ABCA we find
only three minima in the conduction band and a symmetric valence
band (see Fig.~\ref{3Dabca}(a)), analogously with the ABC trilayer
case. For the ABCA system the indirect gap is smaller than the
direct one at low densities, and they coincide at high densities.

When finishing this paper we came aware of a recent
preprint~\cite{koshino2} on the effect of an electric field on
multilayers of graphene with different stacking. They used the
simplest approximation where only $\gamma_0, \gamma_1\neq0$. They
argued that the inclusion of the other tight-binding parameters do
not affect strongly the band structure and the true gap. However,
our calculations show that the true gap can be changed by 30\%.

\section{Conclusions} \label{sec4}

 The effect of different stacking order
on the electric field induced energy gap of three and four layers
of graphene was investigated. For three- as well as for four-layer
graphene the energy gap strongly depends on the choice of
stacking, and we found that the gap is much larger than for the
previously studied Bernal stacking. We found that the true gap for
the ABC trilayer and the ABCA four layer graphene is comparable
with the corresponding gap for bilayer graphene with Bernal
stacking. The account of the circular asymmetry of the spectrum,
which is a consequence of the trigonal warping, considerably
changes the size of the induced electronic gap for the studied
systems.

\begin{acknowledgments} This work was supported by the Flemish Science Foundation (FWO-Vl), the
``Belgian Science Policy'' IAP-program, and the Brazilian Science
Foundation CNPq. One of us (AAA) was supported by the Belgian
Federal Science Policy Office.
\end{acknowledgments}

\end{document}